\documentclass[12pt]{article}

\usepackage{amssymb}
\let\a=\alpha

\newcommand{\beq}{\begin{equation}}
\newcommand{\eeq}{\end{equation}}
\newcommand{\beqn}{\begin{eqnarray}}
\newcommand{\eeqn}{\end{eqnarray}}

\newcommand{\nn}{\nonumber}

\newcommand{\be}{\begin{equation}}
\newcommand{\ee}{\end{equation}}
\newcommand{\ba}{\begin{eqnarray}}
\newcommand{\ea}{\end{eqnarray}}
\newcommand{\bdm}{\begin{displaymath}}
\newcommand{\edm}{\end{displaymath}}

\def\a{\alpha}

\newcommand{\ie}{{\it i.e.\ }}
\newcommand{\eg}{{\it e.g.\ }}

\DeclareMathAlphabet{\mathpzc}{OT1}{pzc}{m}{it}
\def\bea{\begin{eqnarray}}
\def\eea{\end{eqnarray}}
\def\beas{\begin{eqnarray*}}
\def\eeas{\end{eqnarray*}}
\def\sla{\raise.15ex\hbox{$/$}\kern-.57em}

\def\bea{\begin{eqnarray}}
\def\eea{\end{eqnarray}}
\def\de{\partial}

\def\sla{\raise.15ex\hbox{$/$}\kern-.57em}
\def\ie{{\it i.e.}~}
\def\eg{{\it e.g.}~}
\def\ap{{\alpha^\prime}}

\def\a{\alpha}

\def\cA{{\cal A}}
\def\cB{{\cal B}}
\def\cC{{\cal C}}

\def\cE{{\cal E}}
\def\cF{{\cal F}}
\def\cG{{\cal G}}
\def\cH{{\cal H}}
\def\cI{{\cal I}}
\def\cJ{{\cal J}}

\def\cL{{\cal L}}
\def\cM{{\cal M}}
\def\cN{{\cal N}}

\def\cQ{{\cal Q}}
\def\cR{{\cal R}}

\def\cT{{\cal T}}

\def\cZ{{\cal Z}}

\begin{document}
\begin{titlepage}
\begin{flushright}
{ROM2F/2009/19}\\
\end{flushright}
%
%
\begin{center}

{\Large\bf On $\cR^4$ terms and MHV amplitudes \\
in $\cN = 5,6$ supergravity vacua of Type II superstrings}\\

\end{center}
\begin{center}
{\bf Massimo Bianchi}\\
{\sl Dipartimento di Fisica, Universit\`a di Roma ``Tor Vergata''\\
 I.N.F.N. Sezione di Roma ``Tor Vergata''\\
Via della Ricerca Scientifica, 00133 Roma, Italy}\\
and \\
 {\sl Physics Department, Theory Unit, CERN \\ CH 1211,
Geneva 23, Switzerland }
\end{center}
\vskip 1.0cm
\begin{center}
{\large \bf Abstract}
\end{center}

We compute one-loop threshold corrections to $\cR^4$ terms in $\cN
= 5, 6$ supergravity vacua of Type II superstrings. We then
discuss non-perturbative corrections generated by asymmetric
D-brane instantons. Finally we derive generating functions for MHV
amplitudes at tree level in $\cN = 5, 6$ supergravities.



\vfill

\end{titlepage}

\section*{Introduction}

$\cN = 5,6$ supergravities in $D=4$ enjoy many of the remarkable
properties of $\cN = 8$ supergravity. Their massless spectra are
unique and consist solely of the supergravity multiplets. Their
R-symmetries are not anomalous \cite{Marcus:1985yy}. Regular BH
solutions can be found whereby the scalars are stabilized at the
horizon by the attractor mechanism\footnote{For a recent review
see \eg \cite{Ceresole:2010hq}.}. It is thus tempting to
conjecture that if pure $\cN=8$ supergravity turned out to be UV
finite \cite{Bern:2006kd, Bern:2009kd, Stelle:2009zz,
Beisert:2010jx, Kallosh:2010kk} then $\cN = 5,6$ supergravities
should be so, too.

As shown in \cite{Green:2007zzb, Bianchi:2009wj, Bianchi:2009mj},
Type II superstrings or M-theory accommodate  $\cN = 8$
supergravity in such a way as to include non-perturbative states
that correspond to singular BH solutions in $D=4$. The same is
true for $\cN = 5,6$ supergravities. While the embedding of  $\cN
= 8$ supergravity corresponds to simple toroidal
compactifications, the embedding of  $\cN = 5,6$ supergravities,
pioneered by S.~Ferrara and C.~Kounnas in \cite{Ferrara:1989nm}
and recently reviewed in \cite{Bianchi:2008cj}, requires
asymmetric orbifolds \cite{Narain:1986qm, Dabholkar:1998kv} or
free fermion constructions \cite{Antoniadis:1986rn,
Antoniadis:1987wp, Kawai:1986ah, Kawai:1986vd, Kawai:1986va}.

 The inclusion of BPS states, whose possible singular behavior
from a strict 4-d viewpoint is resolved from a higher dimensional
perspective, generate higher derivative corrections to the
low-energy effective action. In particular a celebrated $\cR^4$
term appears that spoils the continuous non-compact symmetry of
`classical' supergravity. Absence of such a term has been recently
shown for pure $\cN=8$ supergravity in \cite{Brodel:2009hu}. In
superstring theory, the $\cR^4$ term receives contribution at
tree-level, one-loop and from non-perturbative effects associated
to D-instantons \cite{Green:1997tv} and other wrapped branes
\cite{Becker:1995kb}. Proposals for the relevant modular form of
the $E_{7(7)}(Z)$ U-duality group have been recently put forward
in \cite{Green:2010wi, Green:2010kv, Pioline:2010kb} that seem to
satisfy all the checks.

In this note we consider one-loop threshold corrections to the
same kind of terms in superstring models with $\cN = 5, 6$
supersymmetry in $D=4$ and $\cN = 6$ in $D=5$. After excluding
$\cR^2$ terms\footnote{$\cR^3$ terms cannot be supersymmetrized on
shell when all particles are in the supergravity multiplet
\cite{Brodel:2009hu}.}, we will derive formulae for the
`perturbative' threshold corrections. In $D=4$ we will also
discuss other MHV amplitudes\footnote{For a recent review see \eg
\cite{Drummond:2010ep}.} that can be obtained by orbifold
techniques from the generating function of $\cN = 8$ supergravity
amplitudes  \cite{Bianchi:2008pu}.

Aim of the analysis is three-fold. First, we would like to show
that $\cN = 5,6$  supersymmetric models in $D=4$ behave very much
as their common $\cN = 8$ supersymmetric parent. Second, (gauged)
$\cN=5, 6$ supergravities have played a crucial role in the recent
understanding of M2-brane dynamics \cite{Bagger:2006sk,
Gustavsson:2007vu, Aharony:2008gk, Aharony:2008ug} and
non-perturbative tests may be refined by considering the effects
of world-sheet instantons in $CP^3$  \cite{Hosomichi:2008ip,
Cagnazzo:2009zh, Bianchi:2010mg, Drukker:2010nc} along the lines
of our present (un-gauged) analysis. Finally, in addition to
world-sheet instantons, D-brane instantons corresponding to
Euclidean bound states of `exotic' D-branes should contribute that
generalize `standard' D-brane instanton calculus to Left-Right
asymmetric backgrounds.

Plan of the note is as follows. In Section \ref{N56sugra}, we
briefly review $\cN = 5,6$ supergravities in $D=4,5$ and their
embedding in Type II superstrings. We then pass to consider in
Section \ref{4grav} a 4-graviton amplitude at one-loop which
allows to derive the `perturbative' threshold corrections to
$\cR^4$ terms, thus excluding $\cR^2$ terms. For simplicity, we
only give the explicit result for $\cN = 6$ in $D=5$ in Section
\ref{N6D5thresh} and sketch how to complete the non-perturbative
analysis by including asymmetric D-brane instantons
\cite{Bianchi:2008cj} in Section \ref{Udual}. Finally, in Section
\ref{MHV} we consider MHV amplitudes in $\cN = 5,6$ supergravities
in $D=4$ and show how they can be obtained at tree-level by
orbifold techniques from the generating function for MHV
amplitudes in $\cN = 8$ supergravity \cite{Bianchi:2008pu}.
Section \ref{Conclus} contains a summary of our results and
directions for further investigation.

\section{Type II Superstring models with $\cN = 5,6$ in $D=4,5$}
\label{N56sugra}
 Let us briefly recall how $\cN = 5,6$
supergravities can be embedded in String Theory. The highest
dimension where classical $\cN = 6$ supergravity  with 24
supercharges can be defined is $D=6$. However the resulting $\cN
=(2,1)$ theory is anomalous and thus inconsistent at the quantum
level \cite{D'Auria:1997cz}. So we are led to consider $D=5$ and
then reduce to $D=4$. $\cN = 5$ supergravity with 20 supercharges
can only be defined is $D=4$ and lower. Although we will only
focus on $\cR^4$ terms in $D=4$ the parent $D=5$ theory is
instrumental to the identification of the relevant BPS instantons.

\subsection{$\cN = 6=2_{_L} + 4_{_R}$ supergravity in $D=5$}

The simplest way to embed $\cN = 6$ in Type II superstrings, is to
quotient a toroidal compactification $T^5 = T^4\times S^1$ by a
chiral $Z_2$ twist of the L-movers (`T-duality') on four internal
directions \be X^i_{_L} \rightarrow - X^i_{_L} \quad , \quad
\Psi^i_{_L} \rightarrow -\Psi^i_{_L} \quad , \quad i=6,7,8,9
\label{Z2twist}\ee accompanied by an order two shift that make
twisted states massive. As a result half of the supersymmetries in
the L-moving sector are broken. The perturbative spectrum is coded
in the one-loop torus partition function.

In the untwisted sector, one finds \be \cT_u = {1\over 2}
\left\{(Q_o + Q_v) \bar{Q} \Lambda_{5,5}[^0_0] + (Q_o - Q_v)(X_o -
X_v) \bar{Q} \Lambda_{1,5}[^0_1] \right\} \ee where $X_o - X_v = 4
\eta^2/\theta_2^2$ (with $\eta$ denoting Dedekind's function and
$\theta_{1,..4}$ denoting Jacobi's elliptic functions) describes
the effect of the $Z_2$ projection on four internal L-moving
bosons, while \bea \Lambda_{l,r}[^a_b] = \sum_{p_{_L}, p_{_R}}
e^{i\pi [a_{_L} p_{_L} - a_{_R} p_{_R}]} q^{{1\over 2} (p_{_L} +
{1\over 2} b_{_L})^2} \bar{q}^{{1\over 2} (p_{_R} + {1\over 2}
b_{_R})^2} \eea are (shifted) Lorentzian lattice sums of signature
$(l,r)$ and $Q=V_8 - S_8$, $Q_o = V_4 O_4 - S_4 S_4$, $Q_v = O_4
V_4 - C_4 C_4$, with $O_n, V_n, S_n, C_n$ the characters of
$SO(n)$ at level $\kappa=1$\footnote{For $n$ odd $S_n$ coincides
with $C_n$ and will be denoted by $\Sigma_n$.}.

At the massless level, in $D=5$ notation with $SO(3)$ little
group, one finds \bea &&(V_3 + O_3 - 2\Sigma_3) \times (\bar{V}_3
+ 5 \bar{O}_3 - 4 \bar{\Sigma}_3)
\rightarrow \\
&&(g + b_2 + \phi)_{_{NS-NS}} + 6 A_{_{NS-NS}} + 5 \phi_{_{NS-NS}}
+ 8 A_{_{R-R}} + 8 \phi_{_{R-R}} - {Fermi} \nn  \eea that form the
$\cN =6$ supergravity multiplet in $D=5$ \be SG^{D=5}_{\cN = 6} =
\{g_{\mu\nu}, 6 \psi_\mu, 15 A_\mu, 20 \chi, 14 \varphi\} \ee The
R-symmetry is $Sp(6)$ while the `hidden' non-compact symmetry is
$SU^*(6)$, of dimension $35$ and rank $3$ generated by $6\times 6$
matrices of the form $Z = (Z_1, Z_2; -\bar{Z}_2,\bar{Z}_1)$ with
$Tr(Z_1 + \bar{Z}_1) = 0$.

For later purposes, let us observe that the 128 massless states of
${\cN = 6}$ supergravity in ${D=5}$ are given by the tensor
product of the 8 massless states of ${\cN = 2}$ SYM (for the
Left-movers) and the 16 massless states of ${\cN = 4}$ SYM (for
the Right-movers) {\it viz.} \be SG^{D=5}_{\cN = 6} =
SYM^{D=5}_{\cN = 2} \otimes SYM^{D=5}_{\cN = 4} = \{A_\mu, 2
\lambda, \phi \}_L \otimes \{\tilde{A}_\nu, 4 \tilde\lambda, 5
\tilde\phi\}_R \ee

After dualizing all masseless 2-forms into vectors, the ${\bf 15}
=7_{_{NS-NS}}+ 8_{_{_{R-R}}}$ vectors transform according to the
antisymmetric tensor of $SU^*(6)$. The
$14=1_{_{NS-NS}}+5_{_{NS-NS}}+8_{_{R-R}}$ scalars parameterize the
moduli space \be \cM^{D=5}_{\cN = 6} = SU^*(6)/Sp(6)\ee

By world-sheet modular transformations (first $S$ and then $T$)
one finds the contribution of the twisted sector \be \cT_t =
{1\over 2} \left\{(Q_s + Q_c)(X_s + X_c) \bar{Q}
\Lambda_{1,5}[^1_0] + (Q_s - Q_c)(X_s - X_c) \bar{Q}
\Lambda_{1,5}[^1_1] \right\}\ee where $X_s + X_c =
4\eta^2/\theta_4^2$, $X_s - X_c = 4\eta^2/\theta_3^2$, $Q_s = O_4
S_4 - C_4 O_4$ (`massless'), $Q_c = V_4 C_4 - S_4 V_4$
(`massive'). Due to the (L-R symmetric) $Z_2$ shift, the massless
spectrum receives no contribution from the twisted sector.
Non-perturbative states associated to L-R asymmetric bound states
of D-branes were studied in \cite{Bianchi:2008cj}. There are
several other ways to embed $\cN = 6$ supergravity in Type II
superstrings, reviewed in \cite{Bianchi:2008cj}.

\subsection{$\cN = 6$ supergravities in $D=4$}

Reducing on another circle with or without further shifts, yields
$\cN =6$ supergravity in $D=4$ \cite{Ferrara:1989nm}.

The massless spectrum is given by \bea &&(V_2 + 2 O_2 - 2S_2 - 2
C_2) \times (\bar{V}_2 + 6 \bar{O}_2 - 4 \bar{S}_2 - 4 \bar{C}_2)
\rightarrow \\
&&(g + b + \phi)_{_{NS-NS}} + 8 A_{_{NS-NS}} + 12 \phi_{_{NS-NS}}
+ 8 A_{_{R-R}} + 16 \phi_{_{R-R}} - {Fermi} \nn  \eea and  gives
rise to the $\cN =6$ supergravity multiplet in $D=4$ \be
SG^{D=4}_{\cN = 6} = \{g_{\mu\nu}, 6 \psi_\mu, 16 A_\mu, 26 \chi,
30 \varphi\} \ee

For later purposes, let us observe that the 128 massless states of
${\cN = 6}$ supergravity in ${D=4}$ are given by the tensor
product of the 8 massless states of ${\cN = 2}$ SYM (for the
Left-movers) and the 16 massless states of ${\cN = 4}$ SYM (for
the Right-movers) {\it viz.} {\it viz.} \be SG^{D=4}_{\cN = 6} =
SYM^{D=4}_{\cN = 2} \otimes SYM^{D=4}_{\cN = 4} = \{A_\mu, 2
\lambda, 2 \phi \}_L \otimes \{\tilde{A}_\nu, 4 \tilde\lambda, 6
\tilde\phi\}_R \ee

The hidden non-compact symmetry is $SO^*(12)$, of dimension $66$
and rank $3$ generated by $12\times 12$ matrices of the form $Z =
(Z_1, Z_2; -\bar{Z}_2,\bar{Z}_1)$ with $Z_1 = - {Z}^t_1$ and $Z_2$
hermitean. They satisfy $L^\dagger \cJ L = \cJ$ with $\cJ = -
\cJ^t = -\cJ^\dagger$ the symplectic metric in 12-d.
 After dualizing
all masseless 2-forms into axions, the
$30=2_{_{NS-NS}}+12_{_{NS-NS}}+16_{_{R-R}}$ scalar parameterize
the moduli space \be \cM^{D=4}_{\cN = 6} = SO^*(12)/U(6) \ee
 The
$16=8_{_{NS-NS}}+8_{_{R-R}}$ vectors together with their magnetic
duals transform according to the ${\bf 32}$ dimensional chiral
spinor representation of $SO^*(12)$.

Due to the (L-R symmetric) $Z_2$ shift, the massless spectrum
receives no contribution from the twisted sector. Non-perturbative
states associated to L-R asymmetric bound states of D-branes were
studied in \cite{Bianchi:2008cj}.

\subsection{$\cN =5=1_{_L} + 4_{_R}$ supergravity in $D=4$}

The highest dimension where $\cN = 5$ supergravity exists is
$D=4$. In $D=5$ because one cannot impose a symplectic Majorana
condition on an odd number of spinors. A simple way to realize
$\cN =5=1_{_L}+4_{_R}$ supergravity in $D=4$ is to combine
$Z^L_2\times Z^L_2$ twists, acting by T-duality along $T^4_{6789}$
and $T^4_{4589}$, with order two shifts, that eliminate massless
twisted states. In \cite{Ferrara:1989nm}, ``minimal'' $\cN = 5$
superstring solutions of this kind have been classified into four
classes which correspond to different choices of the basis sets of
free fermions or inequivalent choices of shifts in the orbifold
language.

Due to the uniqueness of $\cN = 5$ supergravity in $D=4$, all
models display the same massless spectrum \be SG^{D=4}_{\cN = 5} =
\{g_{\mu\nu}, 5 \psi_\mu, 10 A_\mu, 11  \chi, 10  \phi\} \ee

For later purposes, let us observe that the 64 massless states of
${\cN = 5}$ supergravity in ${D=4}$ are given by the tensor
product of the 4 massless states of ${\cN = 1}$ SYM (for the
Left-movers) and the 16 massless states of ${\cN = 4}$ SYM (for
the Right-movers) {\it viz.} \be SG^{D=4}_{\cN = 5} =
SYM^{D=4}_{\cN = 1} \otimes SYM^{D=4}_{\cN = 4} = \{A_\mu, \lambda
\}_L \otimes \{\tilde{A}_\nu, 4 \tilde\lambda, 6 \tilde\phi\}_R
\ee

The massless scalars parameterize the moduli space \be
\cM^{D=4}_{\cN = 5} = SU(5,1)/U(5) \ee The graviphotons together
with their magnetic duals transform according to the ${\bf 20}$
complex (3-index totally antisymmetric tensor) representation of
$SU(5,1)$.

\section{Four-graviton one-loop amplitude}
\label{4grav}

Since $\cN = 5,6$ supergravities can be obtained as asymmetric
orbifolds of  tori, tree-level scattering amplitudes of untwisted
states such as gravitons are identical to the corresponding
amplitudes in the parent $\cN = 8$ theory. In particular, denoting
by $f^{\cN=5,6}_{\cR^4}(\varphi)$ the moduli dependent coefficient
function of the ${\cR^4}$ term, one has \be f^{\cN=5,6}_{\cR^4} =
{2\over n} \zeta(3) {V({\bf T}^d)\over g_s^2 \ell_s^2}  +
{\cI^{\cN=8}_{d,d}\over n \ell_s^2} + ... \ee where $n$ is the
order of the orbifold group, that reduces the volume of ${\bf
T}^d$ with $d=5,6$ to the volume of the orbifold, $\ell_s^2 = \ap$
and $...$ stands for non-perturbative terms. The one-loop
threshold integral is given by \be \cI^{\cN=8}_{d,d} = (2\pi)^d
\int_{\cF} {d^2\tau\over \tau_2^2} \left[
\tau_2^{d/2}\Gamma_{d,d}(G,B;\tau) - \tau_2^{d/2}\right] = 2
\pi^{2-{d\over 2}} \Gamma\left({d\over 2} - 1\right)
\cE^{SO(d,d|Z)}_{{\bf v}={\bf 2d},s={d\over 2} - 1} \ee where \be
\cE^{SO(d,d|Z)}_{{\bf v}={\bf 2d},s={d\over 2} - 1} =
{\sum_{\vec{m},\vec{n}: \vec{m}\cdot\vec{n}=0}} [(\vec{m}+
B\vec{n})^tG^{-1}(\vec{m}+ B\vec{n}) + \vec{n}^tG\vec{n}]^{-d+2}
\ee is a constrained Epstein series that encodes the contribution
of perturbative 1/2 BPS states \ie those satisfying
$\vec{m}\cdot\vec{n}=0$. The subtraction eliminates IR
divergences, \ie the terms with $\vec{m}=\vec{n}=0$. For $\cN =
5,6$ the contribution of the $(r,s)=(0,0)$ `un-twisted' sector is
up to a factor $1/n$ the same as in toroidal Type II
compactifications with restricted metric $G_{ij}$ and
anti-symmetric tensor $B_{ij}$.

In the following we will focus on the contribution of the
`twisted' sectors\footnote{We write `twisted' in quotes, since the
terminology includes projections of the untwisted sector, \ie
amplitudes with $r=0$ and $s=1,...,n-1$} with $(r,s)\neq (0,0)$.

Recall that the partition function reads \be \cZ = \bar\cQ {1\over
n} \sum_{r,s}^{0,n-1} \sum_\a {\theta_\a(0)\over \eta^3}
\prod_{I=1}^3 {\theta_\a(u^I_{rs})\over \theta_1(u^I_{rs})}
\Gamma[^r_s] \ee where $u^I_{rs}$ encode the effect of the
Left-moving twist on the three complex internal directions, while
$\Gamma[^r_s]$ denote the twisted and shifted lattice sums.

Following the analysis in \cite{Bianchi:2006nf} for one-loop
scattering of vector  bosons in unoriented D-brane worlds and
exploiting the `factorization' of world-sheet correlation
functions one has \be \cA_{4h} = {1\over n} \sum_{r,s}^{0,n-1}
\int {d^2\tau\over \tau_2^2}\Gamma[^r_s] \cC_{4v}^L \cC_{4v}^R \ee

Since in both $\cN = 5,6$ cases the orbifold projection only  acts
by a shift of the lattice on the Left-movers, \ie preserves all
four space-time supersymmetries, their contribution is simply \be
\cC_{4v}^L = const \ee after summing over spin structures. In the
terminology of \cite{Bianchi:2006nf} only terms with 4 fermion
pairs contribute. Recall that the graviton vertex in the $q=0$
superghost picture reads \be V_h = h_{\mu\nu} (\de X^\mu + i k\psi
\psi^\mu) (\bar\de\tilde{X}^\nu + i k\tilde\psi \tilde\psi^\nu)
e^{ikX} \ee and, for fixed graviton helicity\footnote{Henceforth
we use $D=4$ notation but the analysis is valid in $D=5$ too.},
one can exploit `factorization' of the physical polarization
tensor \be h^{(2\sigma)}_{\mu\nu} = a_\mu^{(\sigma)}
a_\nu^{(\sigma)} \ee in terms of photon polarization vectors.

In the R-moving sector however, the orbifold projection breaks 1/2
($\cN = 6$) or 3/4 ($\cN = 5$) of the original four space-time
supersymmetries. Correlation functions of two and three fermion
bilinears will be non vanishing, too.

For two fermion bilinears one has \cite{Bianchi:2006nf} \be
\langle\de X^{\mu_1} \de X^{\mu_2} k_3\psi \psi^{\mu_3} k_4\psi
\psi^{\mu_4} \rangle = [\eta^{\mu_1\mu_2}\de_1\de_2 \cG_{12} -
\sum_{i\neq 1} k_i^{\mu_1} \de_1\cG_{1i} \sum_{j\neq 2}
k_j^{\mu_2} \de_2\cG_{1j}] [k_3 k_4 \eta^{\mu_3\mu_4} -
k_3^{\mu_4} k_4^{\mu_3}] \ee where $\cG_{ij}$ denotes the scalar
propagator on the torus (with $\ap =2$) \be \cG_{z,w} = -
\log{|\theta_1(z-w)|\over|\theta'_1(0)|} - \pi {Im(z-w)^2\over
Im\tau} \ee

Similarly, for three fermion bilinears, one finds
\cite{Bianchi:2006nf} \be \langle\de X^{\mu_1} k_2\psi
\psi^{\mu_2} k_3\psi \psi^{\mu_3} k_4\psi \psi^{\mu_4} \rangle =
\sum_{i\neq 1} k_i^{\mu_1} \de_1\cG_{1i} [k_2 k_3k_4^{\mu_2}
\eta^{\mu_3\mu_4} - ...]\omega_{234} \ee with $\omega_{234} =
\de\log\theta_1(z_{23}) +  \de\log\theta_1(z_{34}) +
\de\log\theta_1(z_{42})$

For four fermion bilinears, disconnected contractions yield
\cite{Bianchi:2006nf} \bea &&\langle k_1\psi \psi^{\mu_1} k_2\psi
\psi^{\mu_2} k_3\psi \psi^{\mu_3} k_4\psi \psi^{\mu_4}
\rangle_{disc} =  \{[k_1 k_2 \eta^{\mu_1\mu_2} - k_1^{\mu_2}
k_2^{\mu_1}] [k_3 k_4
\eta^{\mu_3\mu_4} - k_3^{\mu_4} k_4^{\mu_3}]\times \nn \\
&&\: (\wp_{12} + \wp_{34} - \Delta_{rs})  + ...\} \eea where $\wp$
is Weierstrass function
 \bea \wp(z) &&= {1\over z^2} + {\sum_{m,n}}^\prime {1\over (z + n + m\tau)^2}
 - {1\over (n + m\tau)^2} \nn\\
 &&= - \de_z^2 \log\theta_1(z) - 2\eta_1
 = - 2 \de_z^2 \cG(z,\bar{z}) - {\pi^2\over 3} \hat{E}_2 \eea
with $\eta_1 = - \theta_1'''/6\theta_1'$ and $\hat{E}_2$ the
non-holomorphic modular form of weight 2 (Eisenstein series).
Weierstrass function satisfies $\wp(1/2) = e_1$, $\wp(\tau/2) =
e_2$, $ \wp(1/2+\tau/2) = e_3$ with $e_1 + e_2 + e_3 =0 $.

Finally, connected contractions of four fermion bilinears yield
\cite{Bianchi:2006nf} \be \langle k_1\psi \psi^{\mu_1} k_2\psi
\psi^{\mu_2} k_3\psi \psi^{\mu_3} k_4\psi \psi^{\mu_4}
\rangle_{conn} = [k_1^{\mu_4} k_2^{\mu_1} k_3^{\mu_2}k_4^{\mu_3}
\pm ...](\wp_{13} - \omega_{123}\omega_{143} +\Delta_{rs}) \ee
where, for $\cN=6$ \be \Delta_{rs} = \wp(u_{rs}) \ee while, for
$\cN=5$
 \be
\Delta_{rs} = 3\eta_1+ {1\over 6} {\cH'''(u_{rs}) \over
\cH'(u_{rs})} \ee with $\cH'/\cH = \sum_I \de
\log\theta_1(u^I_{rs})$, which is clearly moduli independent,
since no NS-NS moduli survive except for the axio-dilaton.
Dependence on R-R moduli and the axio-dilaton is expected to be
generated by L-R asymmetric bound-states of Euclidean D-branes and
NS5-branes.

\subsection{World-sheet Integrations}

Worldsheet integrations can be performed with the help of $\int
d^2z \de_z\cG_{zw} = 0 = \int d^2z \de^2_z\cG_{zw}$ as well as of
\be \int d^2z d^2w  (\de_z\cG_{zw} )^2 = -\tau_2 \hat{E}_2
{\pi^2\over 3} \ee and \be \int d^2z d^2w
[\eta^{\mu_1\mu_2}\de_1\de_2 \cG_{12} k_1k_2 \cG_{12} -
\sum_{i\neq 1} k_i^{\mu_1} \de_1\cG_{1i} \sum_{j\neq 2}
k_j^{\mu_2} \de_2\cG_{1j}]  = -\tau_2 \hat{E}_2 {\pi^2\over 3}
[\eta^{\mu_1\mu_2}   k_1k_2 -k_1^{\mu_2} k_2^{\mu_1} ] \ee

For $\cN = 6 = 4_L + 2_R$, setting $f^{L/R}_{\mu\nu} = k_\mu
a^{L/R}_\nu - k_\nu a^{L/R}_\mu$, one has \bea &&\cL^{twist}_{eff}
= {1\over n} {\sum_{r,s}}^\prime \int d^2\tau \Gamma [^r_s]
\langle f_1 f_2 f_3 f_4\rangle_L^{MHV} \left\{4 [(f_1f_2)(f_3f_4)
+ ..]_R {\pi^2\over 3} \hat{E}_2 + \right. \\ &&\: \left. +
[(f_1f_2)(f_3f_4) + ..]_R \left(- 2 {\pi^2\over 3} \hat{E}_2 +
\wp(u_{rs})\right) + [(f_1 f_2 f_3 f_4) + ...]_R  \left( - 2
{\pi^2\over 3} \hat{E}_2 - \wp(u_{rs}) \right) \right\} \nn \eea
where, including all permutations, \bea \langle f_1 f_2 f_3
f_4\rangle^{MHV} = (f_1 f_2 f_3 f_4) + (f_1 f_3 f_4 f_2) + (f_1
f_4 f_2 f_3) \nn \\ \qquad - 2(f_1 f_2)( f_3 f_4) - 2(f_1 f_3)
(f_4 f_2) - 2(f_1 f_4)( f_2 f_3) \eea is the structure that
appears in 4-pt vector boson amplitudes, that are necessarily MHV
(Maximally Helicity Violating) in $D=4$\footnote{In $D=5$ there is
more than one `helicity', but the tensor structure has the same
form \cite{Green:1987mn, Green:1987sp}.}.

Combining the R-moving contributions one eventually finds \bea
\cL^{twist}_{eff} = \langle \cR_1 \cR_2 \cR_3 \cR_4\rangle^{MHV}
{1\over n} {\sum_{r,s}}^\prime\int d^2\tau \Gamma [^r_s] \left(+ 2
{\pi^2\over 3} \hat{E}_2 - \wp(u_{rs})\right) \eea where $\cR_i$
denote the linearized Riemann tensors of the four gravitons and
\be  \langle \cR_1 \cR_2 \cR_3 \cR_4\rangle^{MHV} = \langle f_1
f_2 f_3 f_4\rangle_L^{MHV} \langle f_1 f_2 f_3 f_4\rangle_R^{MHV}
\ee reproduces the expected $\cR^4$ structure, which is MHV in
$D=4$, and no lower derivative $\cR^2$ and/or $\cR^3$ terms
\cite{Brodel:2009hu}.

For $\cN = 5 = 4_L + 1_R$ in $D=4$ one gets similar results with
$\cE_{\cN=2_R} = \Gamma [^r_s]$ replaced by $\cE_{\cN=1_R} =
\cI_{ab} \cH'/\cH (\ap\tau_2)^{-2}$ which is moduli independent.

Henceforth we will focus on the  $\cN = 6 = 4_L + 2_R$ case and
explore NS-NS moduli dependence of the one-loop threshold in
$D=5$.

\section{One-loop Threshold Integrals}
\label{N6D5thresh}

One-loop threshold integrals for toroidal compactifications have
been briefly reviewed above and shown to represent the
contribution of the $(r,s)\neq (0,0)$ un-twisted sector. For
$(r,s)\neq (0,0)$ the threshold integrals involve shifted lattice
sums as in heterotic strings with Wilson lines
\cite{Bachas:1997mc, Kiritsis:1997hf, Bianchi:1998vq,
Obers:2001sw, Bianchi:2007rb}.

For simplicity let us discuss here the case of $\cN =6$ in $D=5$.
For definiteness we consider $n=2$ ($Z_2$ shift orbifold) and
start at the special point in the moduli space where ${\bf T}^5 =
{\bf T}^4_{SO(8)} \times {\bf S}^1$. Later on we will include
off-diagonal moduli that effectively behave as Wilson lines.

In the `twisted' $[^0_1]$ sector, the relevant threshold integral
is of the form \bea \cI^{\cN=6}_{1,5}[^0_1] &&= (2\pi)^5 \int_\cF
{d^2\tau \over \tau_2} \tau_2^{5/2}\Gamma_{1,1}[^0_1](R) \bar{O}_8
\left[{2 \pi^2 \over 3} \hat{E}_2 +
\wp(1/2) \right] \nn\\
&& =  (2\pi)^5 \int_\cF {d^2\tau \over \tau_2} \tau_2^{2} {R\over
\sqrt{\ap}} \sum_{m,n} e^{- |2m + (2n+1)\tau|^2 {\pi R^2/
4{\ap}}\tau_2} \bar{O}_8 \left[{2 \pi^2 \over 3} \hat{E}_2 +
\wp(1/2) \right] \eea Setting $(2m, 2n+1) = (2\ell +1) (2m',
2n'+1)$ and using invariance of $\bar{O}_8$ under $\tau\rightarrow
\tau +2$ allows to unfold the integral to the double strip \be
(2\pi)^5 {R\over \sqrt{\ap}} \int_{-1}^{+1} d\tau_1 \int_0^\infty
d\tau_2 \sum_\ell e^{- (2\ell +1)^2 {\pi R^2/ 4{\ap}}\tau_2}
\sum_{N,\bar N}d_{\bar N} \bar{q}^{\bar N} c_N \bar{q}^N \ee where
$\bar{O}_8 = \sum_{N=|\bf{r}|^2/2} d_N \bar{q}^N$ and ${2 \pi^2
\over 3} \hat{E}_2 + \wp(1/2) = \sum_N c_N {q}^N$. Performing the
trivial integral over $\tau_1$ (level matching ${\bar N} = N$) and
the less trivial integral over $\tau_2$ by means of \be
\int_0^\infty dy y^{\nu -1} e^{-c y - b/y} = (b/c)^{\nu/2}
K_\nu(\sqrt{bc})\quad , \ee where $K_\nu(z)$ is a Bessel function
of second kind, finally yields \be
\cI^{\cN=6}_{1,5}[^0_1](R,A_i=0) = (2\pi)^5 \left({R\over
\sqrt{\ap}}\right)^{3/2} \sum_{\ell = 0}^\infty \sum_{N=1}^\infty
(2\ell + 1) \sqrt{N} d_N \sigma_1(N) K_1\left(4\pi(2\ell + 1)
\sqrt{N R\over\sqrt{\ap}}\right)\ee where \be \sigma_1(N) =
\sum_{d|N} = \psi(N)-\psi(1) = {c_N\over N} \ee from the expansion
of $\hat{E}_2$ in powers of $q$.

The result can be easily generalized to the other sectors of the
$Z_2$ orbifold under consideration as well as to different
(orbifold) constructions that give rise to different shifted
lattice sums. Manifest $SO(1,5|Z)$ symmetry can be achieved
turning on off-diagonal components of $B$ and $G$ (subject to
restrictions). Denoting by $2A_i = G_{i5} + B_{i5}$ and observing
that $G_{i5} - B_{i5}=0$ by construction, one finds \bea &&
\cI^{\cN=6}_{1,5}[^r_s](R,A_i) =  (2\pi)^5 {R\over 2 \sqrt{\ap}}
\sum_{\ell=0}^\infty \sum_{\vec{w}\in \Gamma[^r_s]}
c_{\vec{w}^2/2} \int_0^\infty dy (2\ell + 1)e^{- (2\ell +1)^2 {\pi
R^2/ 4{\ap}}y - 2\pi \vec{w}\cdot\vec{w} + 2\pi i
\vec{w}\cdot\vec{A} } \nn  \\
&& =  (2\pi)^5 \left({R\over \sqrt{\ap}}\right)^{3/2}
\sum_{\ell=0}^\infty \sum_{\vec{w}\in \Gamma[^r_s]}
\sigma_1({\vec{w}^2\over 2}) (2\ell + 1) \sqrt{\vec{w}^2\over 2}
e^{2\pi i \vec{r}\cdot\vec{A}} K_1\left(4\pi (2\ell + 1)
\sqrt{\vec{w}^2 R \over 2 \sqrt{\ap}}\right) \nn \\
\eea

Summing up the contributions of the various sectors, \ie various
shifted lattice sums, yields the complete one-loop threshold
correction to the $\cR^4$ terms for $\cN=6$ superstring vacua in
$D=5$. Clearly only NS-NS moduli (except the dilaton) appear that
expose $SO(1,5)$ T-duality symmetry.

The analysis is rather more involved in $D=4$ where one-loop
threshold integrals receive contribution from trivial, degenerate
and non-degenerate orbits \cite{Dixon:1990pc, Mayr:1993kn}.
Alternative methods for unfolding the integrals over the
fundamental domain have been proposed \cite{Trapletti:2002uk,
Cardella:2008nz}.

Explicit computation is beyond the scope of the present
investigation. It proceeds along the lines above and presents
close analogy with threshold computations in $\cN =2$ heterotic
strings sectors in the present of Wilson lines
\cite{Bachas:1997mc, Kiritsis:1997hf, Obers:2001sw} or,
equivalently, $\cN =4$ heterotic strings in $D=8$
\cite{Lerche:1998gz}. Rather than focussing on this interesting
but rather technical aspect of the problem, let us turn our
attention onto the non-perturbative dependence on the other R-R
moduli as well as dilaton. This is brought about by the inclusion
of asymmetric D-brane instantons.

\section{Low-energy action and U-duality}
 \label{Udual}

In \cite{Bianchi:2008cj} the conserved charges coupling to the
surviving R-R and NS-NS graviphotons, were identified as
combinations of those appearing in toroidal compactifications. In
the case of maximal $\cN = 8$ supergravity, the 12 NS-NS
graviphotons couple to windings and KK momenta. Their magnetic
duals to wrapped NS5-branes (H-monoples) and KK monopoles. The 32
R-R graviphotons (including magnetic duals) couple to 6 D1-, 6 D5-
and 20 D3-branes in Type IIB and to 1 D0-, 15 D2- , 15 D4- and 1
D6-branes in Type IIA.

An analogous statement applies to Euclidean branes inducing
instanton effects. In toroidal compactifications with $\cN = 8$
supersymmetry, one has 15 kinds of worldsheet instantons (EF1), 1
D(-1), 15 ED1, 15 ED3 and one each of EN5, ED5, EKK5 for Type IIB.
For Type IIA superstrings one finds 6 ED0, 20 ED2, 6 ED4 and one
each of EN5 and EKK5.

In a series of paper \cite{Green:2010wi, Green:2010kv}, a natural
proposal has been made for the non-perturbative completion of the
modular form of $E_{d+1}(Z)$ that represent the scalar dependence
of the $\cR^4$ and higher derivative terms in $\cN=8$ superstring
vacua. The explicit formulae are rather simple and elegant. In
particular \be f^{\cN=8}_{\cR^4}(\Phi) =
\cE^{E(d+1|Z)}_{[10^d],3/2}(\Phi) \ee where
$\cE^{E(d+1|Z)}_{[10^d],3/2} (\Phi)$ is an Einstein series of the
relevant U-duality group. The above proposal satisfies a number of
consistency checks including perturbative string limit \ie small
string coupling in which $E(d+1|Z) \rightarrow SO(d,d|Z)$ and
$[10..0]\rightarrow {\bf 2d}$), large radius limit in which
$E(d+1|Z) \rightarrow E(d|Z)$ and $[10..0]\rightarrow [10..0]$)
and M-theory limit in which $E(d+1|Z) \rightarrow SL(d+1|Z)$ and
$[10..0]\rightarrow [10..0]'$). Moreover $f_{\cR^4}$ only receives
contribution from 1/2 BPS states as expected for a supersymmetric
invariant that can be written as an integral over half of
(on-shell) superspace.

An independent but not necessarily inequivalent proposal has been
made in \cite{Pioline:2010kb}.

We expect similar results for $\cR^4$ terms in $\cN=5,6$
superstring vacua with the following caveats. First, in $\cN=5,6$
superspace $\cR^4$ terms are 1/5 and 1/3 BPS respectively, since
they require integrations over 16 Grassman variables. Indeed we
have explicitly seen that one-loop threshold correction involve
the left-moving sector, in which supersymmetry is partially
broken, in an essential way. Second, the U-duality group is not of
maximal rank and the same applies to the T-duality subgroup,
present in the $\cN=6$ case. Third, $\cN=5,6$ only exist in $D\le
5$ or $D\le 4$. Some decompactification limits should produce $\cN
=8$ vacua in $D=10$.

Let us try and identify, the relevant 1/3 or 1/5 BPS Euclidean
D-brane bound states.

\subsection{$\cN=6$ ED-branes}

In the Type IIB description, the chiral $Z_2$ projection
(`T-duality') from $\cN = 8$ to $\cN=6$ yields Euclidean D-brane
bound states of the form \bea D(-1)+ED3_{\hat{T}^4} \qquad
ED1_{T^2} + ED5_{T^2\times \hat{T}^4} \qquad ED1_{S^1\times
\hat{S}^1}
+ ED3_{S^1\times \hat{T}^3_\perp} \\
ED1_{\hat{T}^2} + ED1_{\hat{T}^2_\perp} \qquad
ED3_{T^2\times\hat{T}^2} + ED3_{T^2\times\hat{T}^2_\perp}\eea

The above bound states of Euclidean D-branes  are 1/3 BPS since
they preserve 8 supercharges out of the 24 supercharges present in
the background.

A similar analysis applies to world-sheet and ENS5 instantons.

There are several other superstring realizations of $\cN = 6$
supergravity in $D=4$. Given the uniqueness of the low-energy
theory, they all share the same massless spectrum but the massive
spectrum and the relevant (Euclidean) D-brane bound-states depend
on the choice of model.

\subsection{Non-perturbative threshold corrections}

By analogy with $\cN = 8$ one would expect $f_{\cR^4} =
\Theta_{G}$ \ie an automorphic form of the U-duality group $G$ \ie
$G=SO^*(12)$ ($SU^*(6)$) for $\cN=6$ in $D=4$ ($D=5$) and
$G=SU(5,1)$ for $\cN =5$ in $D=4$. The relevant `instantons'
should be associated to BPS particles in one higher dimension
(when possible).

For $\cN=6$, in the decompactification limit the relevant
decomposition under $SO^*(12)\rightarrow SU(5,1)\times R^+$ is \be
{\bf 66} \rightarrow {\bf 35}_0 + {\bf 1}_0 + {\bf 15}_{+2} + {\bf
15'}_{-2} \ee so that the 15 particle charges in $D=6$ satisfy 15
1/3 BPS `purity' conditions in $D=5$ \be {\de \cI_3\over \de
\cQ^{[ij]}}= 0 \ee where $\cI_3^{\cN=6,D=5} =
\varepsilon_{ijklmn}\cQ^{[ij]}\cQ^{[kl]}\cQ^{[mn]}$. The moduli
space decomposes according to \be {SO^*(12)\over U(6)}\supset
{SU(5,1)\over Sp(6)}\times R^{15} \times R^+ \ee More precisely
the 15 charges decompose under $SO(1,5)$ into a 15-dim irrep. The
`purity' conditions include $detQ=0$, viewed as a $6\times 6$
antisymmetric matrix.

For $\cN=6$, in the string theory limit the relevant decomposition
under $SO^*(12)\rightarrow SO(2,6)\times SL(2)_S$ is \be {\bf 32}
\rightarrow ({\bf 8}_v,{\bf 2})_{NS-NS} + ({\bf 8}_s,{\bf
1})_{R-R}+ ({\bf 8}_c,{\bf 1})_{R-R}\ee that yields \be {\bf 66}
\rightarrow ({\bf 28}, {\bf 1}) + ({\bf 1}, {\bf 3}) + ({\bf 8}_s,
{\bf 2}) + ({\bf 8}_c, {\bf 2}) + 3 ({\bf 1}, {\bf 1})  \ee The
moduli space decomposes according to \be {SO^*(12)\over
U(6)}\supset {SO(6,2)\over SO(6)\times SO(2)}\times {SL(2)\over
U(1)} \times R^{16} \ee Further decomposition under $SL(2)_T\times
SL(2)_U \times SL(2)_S$ should allow to get the `non-Abelian' part
of the automorphic from from the `Abelian' one by means of
$SL(2)_{U=\tau} \equiv SL(2)_B$. In particular the action for a
(T-duality invariant) bound state of ED5 and three ED1's into the
action of EN5 and EF1's. While the action of (T-duality invariant)
bound state of ED(-1) and three ED3's is invariant (singlet).
Clearly further detailed analysis is necessary.

\subsection{$\cN=5$ ED-branes}

In the Type IIB description, the two chiral $Z_2$ projections
(`T-duality' on $T^4_{1234}$ and $T^4_{3456}$) from $\cN = 8$ to
$\cN=5$ yield Euclidean D-brane bound states of the form \bea
D(-1)+ED3_{\hat{T}_{1234}^4} + ED3_{\hat{T}_{3456}^4} +
ED3_{\hat{T}_{1256}^4} \\
ED(-1)_{12} +ED5_{123456} + ED1_{34} + ED1_{56} \\
ED1_{i_1i_2} +ED3_{i_1j_2k_3l_3} + ED3_{j_1i_2k_3l_3} + ED1_{j_1j_2}\\
ED1_{i_1i_3} +ED3_{i_1j_2k_2l_3} + ED3_{j_1j_2k_2k_3} +
ED1_{j_1k_3}\\
ED1_{i_2i_3} +ED1_{j_2j_3} + ED3_{i_1j_1j_2i_3} +
ED3_{i_1j_1i_2j_3}\eea

Bound states of Euclidean D-branes carrying the above charges are
1/5 BPS since they preserve 4 supercharges out of the 20
supercharges present in the background.

As in the $\cN = 6$ case, a different analysis applies to BPS
states carrying KK momenta or windings or their magnetic duals.
However, at variant with the $\cN = 6$, the three massive
gravitini cannot form a single complex 2/5 BPS multiplet. One of
them, together with its superpartners, should combined with string
states which are degenerate in mass at the special rational point
in the moduli space where the chiral $Z_2\times Z_2$ projection is
allowed.

\section{Generating MHV amplitudes in $\cN = 5,6$ SG's}
 \label{MHV}

Very much like, tree-level amplitudes in $\cN = 8$ supergravity in
$D=4$ can be identified with `squares' of tree-level amplitudes in
$\cN = 4$ SYM theory \cite{Bern:2006kd, Bern:2009kd}, tree-level
amplitudes in $\cN = 5, 6$ supergravity in $D=4$ can be identified
with `products' of tree-level amplitudes in $\cN = 4$ and  $\cN =
1,2$ SYM theory.

As previously observed, a first step in this direction is to show
that the spectra of $\cN = 5, 6$ supergravity are simply the
tensor products of the spectra of $\cN = 4$ and  $\cN = 1,2$ SYM
theory.

The second step is to work in the helicity basis and focus on MHV
amplitudes\footnote{For a recent review see \eg
\cite{Drummond:2010ep}.}. In $\cN = 4$ SYM the generating function
for (colour-ordered) $n$-point MHV amplitudes is given by
\cite{Nair:1988bq} \be \cF^{{\cN}=4\, SYM}_{MHV} (\eta_i^a,
u_i^\alpha) = {\delta^{8} (\sum_i \eta_i^{a} u_i^\alpha) \over
\langle u_1 u_2\rangle \langle u_2 u_3\rangle ... \langle u_n
u_1\rangle} \ee where $\eta_i^a$ with $i=1,...n$ and $a=1,..4$ are
auxiliary Grassmann variables and $u_i$ are commuting left-handed
spinors, such that $p_i = u_i \bar{u}_i$.

Individual amplitudes obtain by taking derivatives wrt the
Grassman variables $\eta$'s according to the rules \be A^+
\rightarrow 1 \quad \lambda_a^+ \rightarrow {\de \over \de\eta^a}
\quad ... \quad A^- \rightarrow {1\over 4!}
\varepsilon^{abcd}{\de^4\over \de\eta^a .. \de\eta^d} \ee The
intermediate derivatives representing scalars ($\varphi \sim
\de^2/\de\eta^2$) and right-handed gaugini ($\lambda^-\sim
\de^3/\de\eta^3$).

One can reconstruct all tree-level amplitudes, be they MHV or not,
from MHV amplitudes using factorization, recursion relations or
otherwise, see \eg \cite{Drummond:2010ep}.

One can easily derive (super)gravity MHV amplitudes by simply
taking the product of the generating functions for SYM amplitudes
\be \cG^{{\cN}=8 \, SG}_{MHV} (\eta_i^A, u_i^\alpha) = {\cC(u_i)
\delta^{16} (\sum_i \eta_i^{A} u_i^\alpha) \over \langle u_1
u_2\rangle^2 \langle u_2 u_3\rangle^2 ... \langle u_n
u_1\rangle^2} = \cC(u_i) \cF^{\cN=4\,SYM}_{MHV, L} (\eta_i^{a_L},
u_i^\alpha) \cF^{\cN=4\,SYM}_{MHV, R}(\eta_i^{a_R}, u_i^\alpha)
\ee where $\eta^A = (\eta_i^{a_L} , \eta_i^{a_R})$ with $A=1,...8$
and the correction factor $\cC(u_i)$ is only a function of the
spinors $u_i$, actually of the massless momenta $p_i = u_i
\bar{u}_i$ \cite{Bianchi:2008pu}.

The relevant dictionary would read \be h^+ \rightarrow 1 \quad
\psi^+_A \rightarrow {\de \over \de\eta^A} \quad ... \quad h^-
\rightarrow {1\over {\cN}!} {\de^8\over \de\eta^8} \ee

In principle one can reconstruct all tree-level amplitudes, be
they MHV or not, from MHV amplitudes using factorization,
recursion relations or otherwise, see \eg \cite{Drummond:2010ep}.
Unitary methods allow to extend the analysis beyond tree-level. If
all $\cN=8$ supergravity amplitudes were expressible in terms of
squares of $\cN=4$ SYM amplitudes, UV finiteness of the latter
would imply UV finiteness of the former. Although support to this
conjecture at the level of 4-graviton amplitudes, which are
necessarily MHV, seems to exclude the presence of $\cR^4$
corrections, which are 1/2 BPS saturated, it would be crucial to
explicitly test the absence $D^8 \cR^4$ corrections, the first
that are not BPS saturated.

Going back to the problem of expressing MHV amplitudes in
$\cN=5,6$ supergravities in terms of SYM amplitudes, one has to
resort to `orbifold' techniques.

In the $\cN=6$ case, half of the 4 $\eta$'s (say $\eta^3_L$ and
$\eta^4_L$) of the `left' $\cN=4$ SYM factor are to be projected
out \ie `odd' under a $Z_2$ involution. As a result the generating
function is the same as in $\cN=8$ supergravity but the dictionary
gets reduced to \bea && h^+ \rightarrow 1 \quad \psi^+_{A'}
\rightarrow {\de \over \de\eta^{A'}} \quad A^+_{0} = {\de^2 \over
\de\eta_L^{3} \de\eta_L^{4}} \, , \, A^+_{A'B'} = {\de^2 \over
\de\eta^{A'} \de\eta^{B'}} \quad ... \nn
\\ && \quad h^- = {1\over 6!} \varepsilon^{A'_1...A'_6}{\de^{2+6}
\over \de\eta_L^{3} \de\eta_L^{4} \de\eta^{A'_1}...\de\eta^{A'_6}}
\eea where $A'=1,..6$.

Further reduction is necessary for $\cN=5$ case, 3 of the 4
$\eta$'s of the `left' $\cN=4$ SYM factor are to be projected out.
For instance they may acquire a phase $\omega= \exp(i 2\pi/3)$
under a $Z_3$ projection.

The same projections should be implemented on the intermediate
states flowing around the loops. Although tree-level amplitudes in
$\cN=5,6$ supergravity are simply a subset of the ones in $\cN=8$
supergravity, naive extension of the argument at loop order does
not immediately work \cite{Katsaroumpas:2009iy, Bern:2010fy,
Drummond:2010fp}. Several cancellations are not expected to take
place despite the residual supersymmetry of the left SYM factor.
However, in view of the recent observations on the factorization
of $\cN=4$ SYM into a kinematical part and a group theory part,
where the latter satisfies identities similar to the former
\cite{Tye:2010kg, Tye:2010dd, Bern:2010yg} and can thus be
consistently replaced with the former giving rise to consistent
and UV finite $\cN=8$ SG amplitudes, it may well be the case that
a similar decomposition can be used to produce, possibly UV
finite, $\cN=5,6$ SG amplitudes. Our results on $R^4$ lend some
support to this viewpoint.

\section{Conclusions}
 \label{Conclus}

Let us summarize our results. We have shown that the first higher
derivative corrections to the low-energy effective action around
superstring vacua with $\cN=5,6$ supersymmetry are $\cR^4$ terms
as in $\cN =8$. Contrary to $\cN\le 4$, no $\cR^2$ terms appear.
Relying on previous results on vector boson scattering at one-loop
in unoriented D-brane worlds \cite{Bianchi:2006nf}, we have
studied four graviton scattering amplitudes and derived explicit
formulae for the one-loop threshold corrections in asymmetric
orbifolds that realize the above vacua. In addition to a term $1/n
f^{\cN=8}_{cR^4}$, coming from the $(0,0)$ sector, contributions
from non-trivial sectors of the orbifold to $f^{\cN=5,6}_{cR^4}$
display a close similarity with Heterotic threshold corrections in
the presence of Wilson lines \cite{Bachas:1997mc, Kiritsis:1997hf,
Obers:2001sw}. For illustrative purposes, we have computed the
relevant integrals for $\cN=6$ in $D=5$ exposing the expected
$SO(1,5)$ T-duality symmetry. The analysis in $D=4$ is technically
more involved and will be performed elsewhere. We have also
identified the relevant 1/3 or 1/5 BPS bound states of Euclidean
D-branes that contribute to the non-perturbative dependence of the
thresholds on R-R scalars and on the axio-dilaton. By analogy with
$\cN=8$ it is natural to conjecture the possible structure of the
automorphic form of the relevant U-duality group. A more detailed
analysis of this issue is however necessary. Finally, in view of
the potential UV finiteness of $\cN=5,6$ supergravities, we have
discussed how to compute tree-level MHV amplitudes using
generating function and orbifolds techniques
\cite{Bianchi:2008pu}. All other tree-level amplitudes should
follow from factorization and in fact should coincide with $\cN=8$
amplitudes involving only $\cN=5$ or$\cN=6$ supergravity states in
the external legs. Loop amplitudes require a separate
investigation. In particular no generalization of the KLT
relations is known beyond tree level \cite{Kawai:1985xq}.

 \vskip 1cm

 \section*{Acknowledgments}
Discussions with C.~Bachas, M.~Cardella, S.~Ferrara, F.~Fucito,
M.~Green, E.~Kiritsis, S.~Kovacs, L.~Lopez, J.~F.~Morales,
N.~Obers, R.~Poghossyan, R.~Richter, M.~Samsonyan, and
A.~V.~Santini are kindly acknowledged. This work was partially
supported by the ERC Advanced Grant n.226455 {\it ``Superfields''}
and by the Italian MIUR-PRIN contract 2007-5ATT78 {\it
``Symmetries of the Universe and of the Fundamental
Interactions''}.

\end{document}